\documentclass[12pt]{iopart}
\usepackage{amsfonts}
\usepackage{amssymb}
\usepackage{graphicx}
\usepackage{bm}
\usepackage{color}
\usepackage{cite}

\begin{document}

\title{Symmetry operators of the asymmetric two-photon quantum Rabi model}
\author{You-Fei Xie$^{1,*}$ and Qing-Hu Chen$^{1,2,\dag}$}

\address{$^1$ Zhejiang Province Key Laboratory of Quantum Technology and
Device, School of Physics, Zhejiang University, Hangzhou 310027, China}
\address{$^2$ Collaborative Innovation Center of Advanced
Microstructures, Nanjing University, Nanjing 210093, China} %
\eads{\mailto{xieyoufei@zju.edu.cn} and \mailto{qhchen@zju.edu.cn}}
\date{\today }

\begin{abstract}
The true level crossings in a subspace of the asymmetric two-photon quantum
Rabi model (tpQRM) have been observed when the bias parameter of qubit is an
even multiple of the renormalized cavity frequency. Generally, such level
crossings imply the existence of the hidden symmetry because the bias term
breaks the obvious symmetry exactly. In this work, we propose a Bogoliubov
operator approach (BOA) for the asymmetric tpQRM to derive the symmetry
operators associated with the hidden symmetry hierarchically. The explicit
symmetry operators consisting of Lie algebra at low biases can be easily
obtained in our general scheme. We believe the present approach can be
extended for other asymmetric Rabi models to find the relevant hidden
symmetry.
\end{abstract}

\vspace{2pc} \noindent\textit{Keywords\/}: {asymmetric two-photon quantum
Rabi model, hidden symmetry, Bogoliubov operator approach, symmetry operators%
}
\maketitle

\section{Introduction}

The quantum Rabi model (QRM) \cite{Rabi} which describes the simplest
interaction between a two-level system (qubit) and a single-mode cavity is a
paradigmatic model in the quantum optics \cite{book}. In the past decades,
fascinating and inspiring research has proved on the QRM and other related
generalized models, which has been highlighted in \cite%
{reviewJPA,Braak2,Boite,Choi}. Besides, the exact solution of the QRM, which
was firstly found by Braak in the Bargmann space representation \cite{Braak}
and later by Chen \textit{et al.} \cite{Chen2012} in a more familiar Hilbert
space using Bogoliubov operator approach (BOA), further promotes the
developments in the models. On the other hand, the QRM is ubiquitous in the
modern advanced solid devices, such as circuit quantum electrodynamics (QED)
system \cite{Niemczyk,Forn2,Alex}, trapped ions \cite{Wineland}, and quantum
dots \cite{Hennessy}, which can be described in the framework of an
artificial qubit and a resonator coupling system \cite%
{Forn1,Yoshihara,Forn3,Nori}.

In contrast to the cavity QED systems, the static bias of the qubit plays an
essential role in the modern solid devices, resulting in the so-called
asymmetric quantum Rabi model (AQRM). In the superconducting
qubit-oscillator system \cite{Forn1,Yoshihara}, the bias could be
manipulated by the external magnetic flux and the persistent current in the
qubit loop. Recently, the AQRM has attracted a lot of attentions, since the
level crossings in the spectrum have been surprisingly observed in this
model \cite{Zhong,bat,Batchelor,Wakayama} for certain bias values. In
general, the additional bias term breaks the well known $\mathbb{Z}_{2}$%
-symmetry associated with the level crossings, the AQRM should not possess
any obvious symmetry. Nevertheless, it exhibits the phenomenon of energy
level crossings, i.e. double degeneracy \cite{MAJ,lizimin1,lizimin3},
certainly due to the hidden symmetry beyond any known symmetry. To explore
the hidden symmetry, a numerical study \cite{ash2020} was proposed that the
symmetry operator commuting with the Hamiltonian must depend on the
different system parameters unlike the symmetric case. It was quickly
verified analytically in the literature \cite{man,rey1,rey2} that the
symmetry operators can be constructed one by one within Fock space. Most
recently, the present authors also reproduced the symmetry operators for
arbitrary integer biases within BOA in a more simple and general scheme \cite%
{xie2022}. In addition, the symmetry operators of other related asymmetric
one-photon models have been investigated in the references \cite%
{Xilin1,Xilin2}.

Besides the proposals and experimental realizations of the various QRM
models, the two-photon quantum Rabi model (tpQRM) also has been realized in
superconducting circuits \cite{Bertet,Felicetti}, or simulated in trapped
ions \cite{Felicetti15,Puebla} and cold atoms \cite{Schneeweiss} to explore
new quantum effects. In particular, the asymmetric tpQRM could be realized
by the addition of an external bias current \cite{Felicetti,Mooij}, namely
\begin{equation}
H_{tp}=\frac{\Delta }{2}\sigma _{z}+\frac{\epsilon }{2}\sigma _{x}+\omega
a^{\dag }a+g\left[ \left( a^{\dag }\right) ^{2}+a^{2}\right] \sigma _{x},
\label{Hamiltonian}
\end{equation}%
where the first two terms fully describe a qubit with the energy splitting $%
\Delta $ and the static bias $\epsilon$, $\sigma _{x,z}$ are the Pauli
matrices, $a^{\dag }$ and $a$ are the creation and annihilation operators
with the cavity frequency $\omega $, and $g$ is the qubit-cavity coupling
strength.

For the symmetric case ($\epsilon =0$), the exact solution of the tpQRM was
first discovered by Chen \textit{et al.} \cite{Chen2012} using BOA in 2012.
Very interestingly, Braak reproduced this solution in the Bargmann space
\cite{Braaktp} after 10 years, and pointed out that only the $G$-function
proposed in \cite{Chen2012} exhibits an explicitly known pole structure
which dictates the approach to the collapse point \cite{duan2016}. The novel
behavior of the spectral collapse and dynamics for the tpQRM have attracted
increasing attentions \cite{Ng,Felicetti15,Cong,fanheng,zhiguo,Chan,Lo}. It
has been investigated that the tpQRM has $\mathbb{Z}_{4}$-symmetry generated
by the parity $\hat{P}=\exp (i\pi a^{\dag }a/2)\sigma _{z}$. The eigenvalues
$\pm 1$, $\pm i$ of $\hat{P}$ are always independent of the system
parameters, and thus the whole Hilbert space $\mathcal{H}$ separates into
four invariant subspaces. The parity $\hat{P}^{2}=\exp (i\pi a^{\dag }a)$
associated with $\mathbb{Z}_{2}$-symmetry acts only in the bosonic part of $%
\mathcal{H}$ with eigenvalues $\pm 1$ corresponding to the even ($\mathcal{H}%
_{q=1/4}$) and odd ($\mathcal{H}_{q=3/4}$) bosonic subspaces \cite%
{duan2016,Braaktp}. In the same $q$-subspace, there are degenerate level
crossings called Juddian solutions in the spectrum, which is related to the
symmetry group $\mathbb{Z}_{2}\sim $ $\mathbb{Z}_{4}/\mathbb{Z}_{2}$
corresponding to the well known parity of the one-photon QRM \cite{Braak}.

While for the asymmetric case ($\epsilon \neq 0$), the presence of the bias
term breaks the level crossings in the same $q$-subspace. As a result, the $%
\mathbb{Z}_{2}$-symmetry is broken in the subspace. Recently, the level
crossings in the same $q$-subspace of the asymmetric tpQRM have been also
uncovered by the present authors~\cite{xie2021} if the bias value is an even
multiple of the renormalized cavity frequency ($\beta =\sqrt{\omega
^{2}-4g^{2}}$). These level crossings determined by constraint polynomials
are indeed double degenerate, instead of the avoid level crossings.
Therefore, for the asymmetric tpQRM, the hidden symmetry associated with
symmetry operators deserves further mathematical and physical exploration.

In this paper, we derive the symmetry operators of the asymmetric tpQRM with
a BOA scheme. The paper is structured as follows: In section 2, we apply the
BOA and $su(1,1)$ Lie algebra for the asymmetric tpQRM to derive the
constraint relations for the symmetry operators. In section 3, the solutions
of the symmetry operators for the asymmetric tpQRM at low biases are
calculated specifically. Based on the symmetry operators, the
characteristics of level crossings are discussed. A brief summary is given
in the last section.

\section{General scheme within BOA}

To facilitate the BOA scheme, we rewrite the Hamiltonian after a unitary
transformation $\exp (\frac{i\pi }{4}\sigma _{y})$ for the asymmetric tpQRM (%
\ref{Hamiltonian}) in the matrix form (unit is taken of $\omega =1$)
\begin{equation}
H_{0}=\left(
\begin{array}{ll}
a^{\dagger }a+g\left[ \left( a^{\dagger }\right) ^{2}+a^{2}\right] +\frac{%
\epsilon }{2} & ~~~~~~~~-\frac{\Delta }{2} \\
~~~~~~~~-\frac{\Delta }{2} & a^{\dagger }a-g\left[ \left( a^{\dagger
}\right) ^{2}+a^{2}\right] -\frac{\epsilon }{2}%
\end{array}%
\right).
\end{equation}
The level crossings in $\mathcal{H}_{1/4}(\mathcal{H}_{3/4})$ of the
asymmetric tpQRM has been demonstrated in \cite{xie2021} if the bias
parameter of qubit is an even multiple of the renormalized cavity frequency,
i.e. $\epsilon /(2\beta )=N$, $N$ is an integer. To discuss the hidden
symmetry responsible for this double degeneracy, we will follow the same BOA
scheme in \cite{xie2022}.

To generate a simple quadratic form of the diagonal Hamiltonian matrix
elements, we first perform the following Bogoliubov transformation on the
bosonic operators,%
\begin{equation}
a_{+}=ua+va^{\dagger },\qquad a_{+}^{\dagger }=ua^{\dagger }+va,
\label{transform}
\end{equation}%
with%
\[
u=\sqrt{\frac{1+\beta }{2\beta }},\qquad v=\sqrt{\frac{1-\beta }{2\beta }},
\]%
and $\beta =\sqrt{1-4g^{2}}$ whereby $\left[ a_{+},a_{+}^{\dagger }\right]
=1 $ is still met. Then the upper diagonal element becomes
\[
H_{11}=a^{\dagger }a+g\left[ \left( a^{\dagger }\right) ^{2}+a^{2}\right] +%
\frac{\epsilon }{2}=\beta a_{+}^{\dagger }a_{+}-\frac{1-\beta }{2}+\frac{%
\epsilon }{2}.
\]%
Analogously, another Bogoliubov operators are introduced as
\begin{equation}
a_{-}=ua-va^{\dagger },\qquad a_{-}^{\dagger }=ua^{\dagger }-va,
\end{equation}%
leading to
\[
H_{22}=\beta a_{-}^{\dagger }a_{-}-\frac{1-\beta }{2}-\frac{\epsilon }{2}.
\]

For convenience, we employ the operators in a representation of the
non-compact $su(1,1)$ Lie algebra,
\begin{equation}
K_{0}^{a_{i}}=\frac{1}{2}\left( a_{i}^{\dagger }a_{i}+\frac{1}{2}\right)
,\quad K_{+}^{a_{i}}=\frac{1}{2}\left( a_{i}^{\dagger }\right) ^{2},\quad
K_{-}^{a_{i}}=\frac{1}{2}a_{i}^{2}.
\end{equation}%
The $su(1,1)$ generators obey spin-like commutation relations
\[
\left[ K_{0}^{a_{i}},K_{\pm }^{a_{i}}\right] =\pm K_{\pm }^{a_{i}},\qquad %
\left[ K_{-}^{a_{i}},K_{+}^{a_{i}}\right] =2K_{0}^{a_{i}},
\]%
where $a_{i}=a$, $a_{+}$ and $a_{-}$, respectively.

In terms of the Lie algebra, the Hamiltonian can be expressed as
\begin{equation}
H=\left(
\begin{array}{ll}
2\beta K_{0}^{a_{+}}-\frac{1}{2}+\frac{\epsilon }{2} & ~~~~~~-\frac{\Delta }{%
2} \\
~~~~~~-\frac{\Delta }{2} & 2\beta K_{0}^{a_{-}}-\frac{1}{2}-\frac{\epsilon }{%
2}%
\end{array}%
\right) .  \label{Ht}
\end{equation}
Given the diagonalized Hamiltonian, we move to the symmetry operator $J$
associated with the hidden symmetry. It should satisfy the commutation
condition $\left[ J,H\right] =0$, as same as the one-photon AQRM in \cite%
{man,rey1,rey2,xie2022}. For the asymmetric tpQRM, we define $J$ as
\begin{equation}
J=e^{i\pi \frac{a^{\dagger }a}{2}}Q.
\end{equation}%
Note that the form of $J$ is not necessarily unique, here we only show one
concise form for it. By the relations $e^{\frac{i\pi a^{\dag }a}{2}}a=-iae^{%
\frac{i\pi a^{\dag }a}{2}}$, $e^{\frac{i\pi a^{\dag }a}{2}}a^{\dag
}=ia^{\dag }e^{\frac{i\pi a^{\dag }a}{2}}$, we have
\begin{equation}
QH=\widetilde{H}Q,  \label{eq_H}
\end{equation}%
where the Hamiltonian $\widetilde{H}$ reads
\[
\widetilde{H}=\left(
\begin{array}{ll}
2\beta K_{0}^{a_{-}}-\frac{1}{2}+\frac{\epsilon }{2} & ~~~~~~-\frac{\Delta }{%
2} \\
~~~~~-\frac{\Delta }{2} & 2\beta K_{0}^{a_{+}}-\frac{1}{2}-\frac{\epsilon }{2%
}%
\end{array}%
\right) .
\]%
Similar to the one-photon AQRM case, the matrix $Q$ is described as
\[
Q=\left(
\begin{array}{cc}
A & B \\
C & D%
\end{array}%
\right) .
\]

Using the relation (\ref{eq_H}), the four elements in the matrix $Q$ result
in the four equations below,
\begin{eqnarray}
AK_{0}^{a_{+}}-K_{0}^{a_{-}}A~+\frac{\Delta }{4\beta }\left( C-B\right) &=&0,
\label{s1} \\
K_{0}^{a_{+}}D-DK_{0}^{a_{-}}+\frac{\Delta }{4\beta }\left( C-B\right) &=&0,
\label{s2} \\
\left[ B,K_{0}^{a_{-}}\right] -\frac{\epsilon }{2\beta }B+\frac{\Delta }{%
4\beta }\left( D-A\right) &=&0,  \label{s3} \\
\left[ K_{0}^{a_{+}},C\right] -\frac{\epsilon }{2\beta }C+\frac{\Delta }{%
4\beta }\left( D-A\right) &=&0.  \label{s4}
\end{eqnarray}%
We will search for all the operators $A,B,C,D$ which satisfy the above
equalities. The last two equations can be further reduced to%
\begin{equation}
\left[ B,K_{0}^{a_{-}}\right] -\frac{\epsilon }{2\beta }B=\left[
K_{0}^{a_{+}},C\right] -\frac{\epsilon }{2\beta }C.  \label{s34}
\end{equation}%
Note the Lie algebra has the general commutation relations,
\begin{equation}
\left[ \left( K_{-}^{a_{-}}\right) ^{N},K_{0}^{a_{-}}\right] =N\left(
K_{-}^{a_{-}}\right) ^{N},\qquad \left[ K_{0}^{a_{+}},\left(
-K_{+}^{a_{+}}\right) ^{N}\right] =N\left( -K_{+}^{a_{+}}\right) ^{N},
\label{Kr}
\end{equation}%
where $N=\epsilon/(2\beta)$ is an integer for the level crossing case.

Comparing above equations (\ref{Kr}) with (\ref{s34}), we speculate the four
elements in the matrix $Q$ are%
\begin{equation}
M=\sum_{n,m=0}^{n+m\leqslant 2N}M_{n,m}\left( K_{+}^{a_{+}}\right) ^{\frac{n%
}{2}}\left( K_{-}^{a_-}\right) ^{\frac{m}{2}},\qquad M=A,B,C,D,
\label{guess}
\end{equation}%
where the coefficient $M_{n,m}$ depends on the parameters $\beta,g,\Delta
,\epsilon $. Here the operator basis $\left( K_{+}^{a_{+}}\right) ^{\frac{n}{%
2}}\left( K_{-}^{a_-}\right) ^{\frac{m}{2}}$ is equivalent to $2^{-\frac{n+m%
}{2}}\left( a_{+}^{\dag }\right) ^{n}\left( a_{-}\right) ^{m}$, while the
commutation relation $[a_{-},a_{+}^{\dag }]=1/\beta $. To reduce the power
series of $M$, we choose the Lie algebra products of $K_{+}^{a_{+}}$ and $%
K_{-}^{a_-}$ instead of the normal Bogoliubov operators $a_{+}^{\dag }$ and $%
a_{-}$ to expand the four elements. It would be also very cumbersome if we
expand the four elements in terms of operators $a$, $a^{\dag }$ in the
original Fock space. The index $n+m$ (the corresponding degree is $(n+m)/2$)
in the general expression (\ref{guess}) is restricted to $2N$. The highest
index more than $2N$ is also allowed, but the operator $M$ will be indeed
complex and lengthy with additional polynomial terms of high degree. We
deduce that $n+m=2N$ is the minimum among the highest indices for the
operator $M$ at present, as long as it satisfy the constraint conditions (%
\ref{s1}--\ref{s4}). It should be noted that $n+m$ must be an even number
due to the remaining $\mathbb{Z}_{2}$-symmetry, which means the operator $M$
must be located in the $q$-subspace. Based on (\ref{guess}), comparing (\ref%
{s34}) with (\ref{Kr}), to eliminate the highest degree $N$-terms, the
highest order in the operators $B$ and $C$ are given by
\begin{equation}
B=\left( K_{-}^{a_{-}}\right) ^{N}+..., \qquad C=\left(
-K_{+}^{a_{+}}\right) ^{N}+....  \label{BC}
\end{equation}

Let us now consider the symmetry operator $J$ is self-adjoint, leading to%
\[
e^{\frac{i\pi a^{\dag }a}{2}}Qe^{-\frac{i\pi a^{\dag }a}{2}}=Q^{\dag },
\]%
and thus the coefficients have
\begin{equation}
\fl A_{n,m}=\left( -1\right) ^{\frac{n+m}{2}}A_{m,n},\quad B_{n,m}=\left(
-1\right) ^{\frac{n+m}{2}}C_{m,n},\quad D_{n,m}=\left( -1\right) ^{\frac{n+m%
}{2}}D_{m,n}.  \label{coeff}
\end{equation}
Inserting (\ref{guess}) to (\ref{s1}--\ref{s4}), using (\ref{coeff}) and
comparing terms in $\left( K_{+}^{a_{+}}\right) ^{\frac{n}{2}}\left(
K_{-}^{a_-}\right) ^{\frac{m}{2}}$ respectively, one can obtain the
following recurrence equations for the coefficients,
\begin{eqnarray}
&&\fl A_{n-2,m}+A_{n,m-2}~+\frac{m+1}{\beta }A_{n-1,m+1}+\frac{m-n}{4g}%
A_{n,m}+\frac{n+1}{\beta }A_{n+1,m-1}  \nonumber \\
&&\fl+\frac{\left( m+1\right) \left( m+2\right) }{4\beta ^{2}}A_{n,m+2}+%
\frac{\left( n+1\right) \left( n+2\right) }{4\beta ^{2}}A_{n+2,m} = \frac{%
\Delta }{8g\beta }\left( B_{n,m}-\left( -1\right) ^{\frac{n+m}{2}%
}B_{m,n}\right) ,  \label{r1} \\
&&\fl D_{n-2,m}+D_{n,m-2}+\frac{n-m}{4g}D_{n,m}=\frac{\Delta }{8g\beta }%
\left( B_{n,m}-\left( -1\right) ^{\frac{n+m}{2}}B_{m,n}\right) ,  \label{r2}
\\
&&\fl\left( \frac{m-n}{2}-N\right) B_{n,m}+\frac{2g}{\beta }\left(
n+1\right) B_{n+1,m-1}+\frac{g}{2\beta ^{2}}\left( n+1\right) \left(
n+2\right) B_{n+2,m}  \nonumber \\
&& \fl = \frac{\Delta }{4\beta }\left( A_{n,m}-D_{n,m}\right) ,  \label{r3}
\\
&&\fl \left( \frac{n-m}{2}-N\right) C_{n,m}-\frac{2g}{\beta }\left(
m+1\right) C_{n-1,m+1}-\frac{g}{2\beta ^{2}}\left( m+1\right) \left(
m+2\right) C_{n,m+2}  \nonumber \\
&& \fl = \frac{\Delta }{4\beta }\left( A_{n,m}-D_{n,m}\right) .  \label{r4}
\end{eqnarray}
The four constraint relations are the crucial ones to determine the symmetry
operator $J$. In order to solve the four elements in (\ref{guess}), we will
discuss the coefficient $M_{n,m}$ in detail one by one. For convenience, we
focus on the derivation of the coefficients $A_{n,m}$, $B_{n,m}$ and $D_{n,m}
$. The coefficient $C_{n,m}$ can be obtained from (\ref{coeff})
straightforwardly.

\section{Symmetry operators}

We derive the symmetry operators of the asymmetric tpQRM within BOA scheme
rigorously in this section. Recall the symmetry operator $J_{N}=e^{\frac{%
i\pi a^{\dag }a}{2}}Q_{N}$ for $N=\epsilon/(2\beta)$. Using the constraint
relations (\ref{r1}--\ref{r4}) of the coefficients in the previous section,
we first demonstrate the symmetry operators for $N=0,1,2,3$ in detail. In
particular, considering (\ref{r1}) and (\ref{r2}) for $n+m=2N+2$, the two
recurrence relations are reduced to%
\[
A_{n-2,m}+A_{n,m-2}~+\frac{m+1}{\beta }A_{n-1,m+1}=0,\qquad
D_{n-2,m}+D_{n,m-2}=0.
\]%
Given the initial values $A_{0,2N}=D_{0,2N}=0$, increasing $n$ one by one
until $2N+2$, we can obtain
\begin{equation}
A_{n,2N-n}=D_{n,2N-n}=0.  \label{AD}
\end{equation}

\textsl{The symmetry operator for $N=0$.} In this case, we can immediately
have $B=C=1$ from (\ref{BC}), then (\ref{AD}) gives $A=D=0$ easily.
Interestingly, the solution is given by
\begin{equation}
J_{0}=e^{\frac{i\pi a^{\dag }a}{2}}\sigma _{x},  \label{J0}
\end{equation}%
which is just the parity operator in the symmetric tpQRM \cite{duan2016}. In
the following, we discuss the symmetry operators in the asymmetric cases.

\textsl{The symmetry operator for $N=1$.} According to (\ref{BC}), we infer
that
\begin{equation}
B=K_{-}^{a_{-}},\qquad C=-K_{+}^{a_{+}},
\end{equation}%
where the coefficient $B_{0,2}=1$. Then we move to the operators $A$ and $D$%
. Equation (\ref{AD}) has given $A_{0,2}=A_{2,0}=0$ and $D_{0,2}=D_{2,0}=0$.
Inserting $B_{0,2}$ to the relations (\ref{r1}) and (\ref{r2}) for $n=0,m=2$
respectively, one can get
\[
A_{0,0}=\frac{\Delta }{8g\beta },\qquad D_{0,0}=\frac{\Delta }{8g\beta }.
\]%
Therefore, the solution is given by
\begin{equation}
J_{1}=e^{\frac{i\pi a^{\dag }a}{2}}\left(
\begin{array}{cc}
\frac{\Delta }{8g\beta } & K_{-}^{a_-} \\
-K_{+}^{a_{+}} & \frac{\Delta }{8g\beta }%
\end{array}%
\right) .  \label{J1}
\end{equation}%
One can see the symmetry operator still have a simple expression for $N=1$,
but for $N=2$ and $3$, the expressions would be slightly more complicated
with the increasing degree.

\textsl{The symmetry operator for $N=2$.} In this case, we assume
\begin{equation}
B=\left( K_{-}^{a_-}\right) ^{2}+B_{0,0},
\end{equation}%
where $B_{0,4}=1$ and $B_{0,0}$ is to be determined. For simplicity, here we
set the coefficients $B_{n,m}$ for $n+m=2$ are equivalent to zeros. Similar
to the above case $N=1$, we can find $A_{n,4-n}=D_{n,4-n}=0$ by (\ref{AD}).
Then for $n=0,m=4$, the relations (\ref{r2}) and (\ref{coeff}) gives
\[
D_{0,2}=-D_{2,0}=\frac{\Delta }{8g\beta },
\]%
for $n=2,m=0$, the relation (\ref{r2}) yields $D_{0,0}+\frac{1}{2g}D_{2,0}=0$%
, where
\[
D_{0,0}=\frac{\Delta }{16g^{2}\beta }.
\]%
On the basis of the above process to derive the operator $D$, we can get the
operator $A$ similarly and results in $A_{2,0}=-A_{0,2}=$ $-\frac{\Delta }{%
8g\beta }, A_{0,0}=-\frac{\Delta }{16g^{2}\beta }.$ Now there only remains $%
B_{0,0}$ to solve, inserting the values of $A_{0,0}$ and $D_{0,0}$ to the
relation (\ref{r3}) for $n=m=0$ results in
\[
B_{0,0}=\frac{\Delta ^{2}}{64g^{2}\beta ^{2}}.
\]%
Using the equation (\ref{coeff}), the operator $C$ can be immediately
arrived. Summarily, the symmetry operator for $N=2$ is given by
\begin{equation}
J_{2}=e^{\frac{i\pi a^{\dag }a}{2}}\left(
\begin{array}{cc}
\frac{\Delta }{8g\beta }\left( -K_{+}^{a_{+}}+K_{-}^{a_-}\right) -\frac{%
\Delta }{16g^{2}\beta } & \left( K_{-}^{a_-}\right) ^{2}+\frac{\Delta ^{2}}{%
64g^{2}\beta ^{2}} \\
\left( K_{+}^{a_{+}}\right) ^{2}+\frac{\Delta ^{2}}{64g^{2}\beta ^{2}} &
\frac{\Delta }{8g\beta }\left( -K_{+}^{a_{+}}+K_{-}^{a_-}\right) +\frac{%
\Delta }{16g^{2}\beta }%
\end{array}%
\right) .  \label{J2}
\end{equation}%
One can find the symmetry operator still keeps a compact way.

\textsl{The symmetry operator for $N=3$.} In this case, we can preliminarily
set
\begin{equation}
B=\left( K_{-}^{a_-}\right)
^{3}+B_{2,0}K_{+}^{a_{+}}+B_{0,2}K_{-}^{a_-}+B_{0,0}.
\end{equation}%
Inserting $n=0$ one by one in the condition of $n+m=6$, by the equations (%
\ref{AD}) and (\ref{r1}) we see
\[
A_{0,4}=A_{4,0}=-A_{2,2}=\frac{\Delta }{8g\beta }.
\]%
By the relation (\ref{r2}), we also have%
\[
D_{0,4}=D_{4,0}=-D_{2,2}=\frac{\Delta }{8g\beta }.
\]%
For $n+m=4$, the relation (\ref{r1}) is reduced to
\[
A_{n-2,m}+A_{n,m-2}~+\frac{m+1}{\beta }A_{n-1,m+1}+\frac{m-n}{4g}A_{n,m}+%
\frac{n+1}{\beta }A_{n+1,m-1}=0.
\]%
Starting from $m=0$, we have $A_{2,0}=\frac{\Delta }{8g^{2}\beta }$, then
increasing $m$ one by one, the coefficients are
\[
A_{1,1}=-\frac{\Delta }{4g\beta ^{2}},\qquad A_{0,2}=-\frac{\Delta }{%
8g^{2}\beta }.
\]%
To consider the operator $D$, for $n+m=4$, the relation (\ref{r2}) becomes
\[
D_{n-2,m}+D_{n,m-2}+\frac{n-m}{4g}D_{n,m}=0.
\]%
When $n=0,2$ respectively, the above relation arrives%
\[
D_{0,2}=-D_{2,0}=\frac{\Delta }{8g^{2}\beta }.
\]%
For $n+m=2$, if $n=0,2$, using the known coefficients, the relation (\ref{r3}%
) gives
\[
B_{0,2}=\frac{\Delta ^{2}}{32g^{2}\beta ^{2}},\qquad B_{2,0}=-\frac{\Delta
^{2}}{64g^{2}\beta ^{2}}.
\]%
Then we refer to the relations (\ref{r1}) and (\ref{r2}), for $n=2,m=0,$
inserting the obtained coefficients above results in%
\[
A_{0,0}=\frac{\Delta }{16g^{3}\beta }+\frac{\Delta ^{3}}{512g^{3}\beta ^{3}}-%
\frac{\Delta }{16g\beta ^{3}},\qquad D_{0,0}=\frac{\Delta }{16g^{3}\beta }+%
\frac{\Delta ^{3}}{512g^{3}\beta ^{3}}.
\]
Therefore, the symmetry operator is shown as $J_{3}=e^{\frac{i\pi a^{\dag }a%
}{2}}\left(
\begin{array}{cc}
A_{3} & B_{3} \\
C_{3} & D_{3}%
\end{array}%
\right) $ with
\begin{eqnarray}
A_{3} &=&\frac{\Delta }{8g\beta }\left[ \left( K_{+}^{a_{+}}\right)
^{2}-K_{+}^{a_{+}}K_{-}^{a_-}+\left( K_{-}^{a_-}\right) ^{2}\right] +\frac{%
\Delta }{8g^{2}\beta }\left( K_{+}^{a_{+}}-K_{-}^{a_-}\right)  \nonumber \\
&&-\frac{\Delta }{4g\beta ^{2}}\left( K_{+}^{a_{+}}\right) ^{\frac{1}{2}%
}\left( K_{-}^{a_-}\right) ^{\frac{1}{2}}+\frac{\Delta }{16g^{3}\beta }+%
\frac{\Delta ^{3}}{512g^{3}\beta ^{3}}-\frac{\Delta }{16g\beta ^{3}},
\nonumber \\
B_{3} &=&\left( K_{-}^{a_-}\right) ^{3}-\frac{\Delta ^{2}}{32g^{2}\beta ^{2}}%
\left( \frac{1}{2}K_{+}^{a_{+}}-K_{-}^{a_-}\right) ,  \nonumber \\
C_{3} &=&\left( -K_{+}^{a_{+}}\right) ^{3}-\frac{\Delta ^{2}}{32g^{2}\beta
^{2}}\left( K_{+}^{a_{+}}-\frac{1}{2}K_{-}^{a_-}\right) ,  \nonumber \\
D_{3} &=&\frac{\Delta }{8g\beta }\left[ \left( K_{+}^{a_{+}}\right)
^{2}-K_{+}^{a_{+}}K_{-}^{a_-}+\left( K_{-}^{a_-}\right) ^{2}\right] -\frac{%
\Delta }{8g^{2}\beta }\left( K_{+}^{a_{+}}-K_{-}^{a_-}\right)  \nonumber \\
&&+\frac{\Delta }{16g^{3}\beta }+\frac{\Delta ^{3}}{512g^{3}\beta ^{3}}.
\label{J3}
\end{eqnarray}

We have checked that all the coefficients of the symmetry operators satisfy
the constraint relations (\ref{r1}--\ref{r4}), and they are the unique
solutions for $N=0,1,2,3$ in the present scheme. In the solutions within BOA
for the symmetry operator, one can see many terms of the same degree share
the same coefficients, such as the dominant terms ($n+m=2N-2$) in $D$ (also
in $A$). The terms for $n+m=2N-2$ in $B$ (also in $C$) even vanish which
simplify the symmetry operators effectively. If the symmetry operator within
BOA is expanded in the original Fock space, for example, the term $\left(
K_{+}^{a_{+}}\right) ^{N}$ $\sim \left( ua^{\dagger }+va\right) ^{2N}$, the
largest degree will rise to $2N$ and the power series will increase rapidly
along with $2N$. In the present work, we focus on the low biases to
demonstrate our BOA scheme. Since the constraint relations (\ref{r1}--\ref%
{r4}) have been given, one can continue to compute the symmetry operator for
larger biases if interested, which should be feasible within our BOA scheme.

\begin{figure}[tbph]
\centering
\includegraphics[width=14cm]{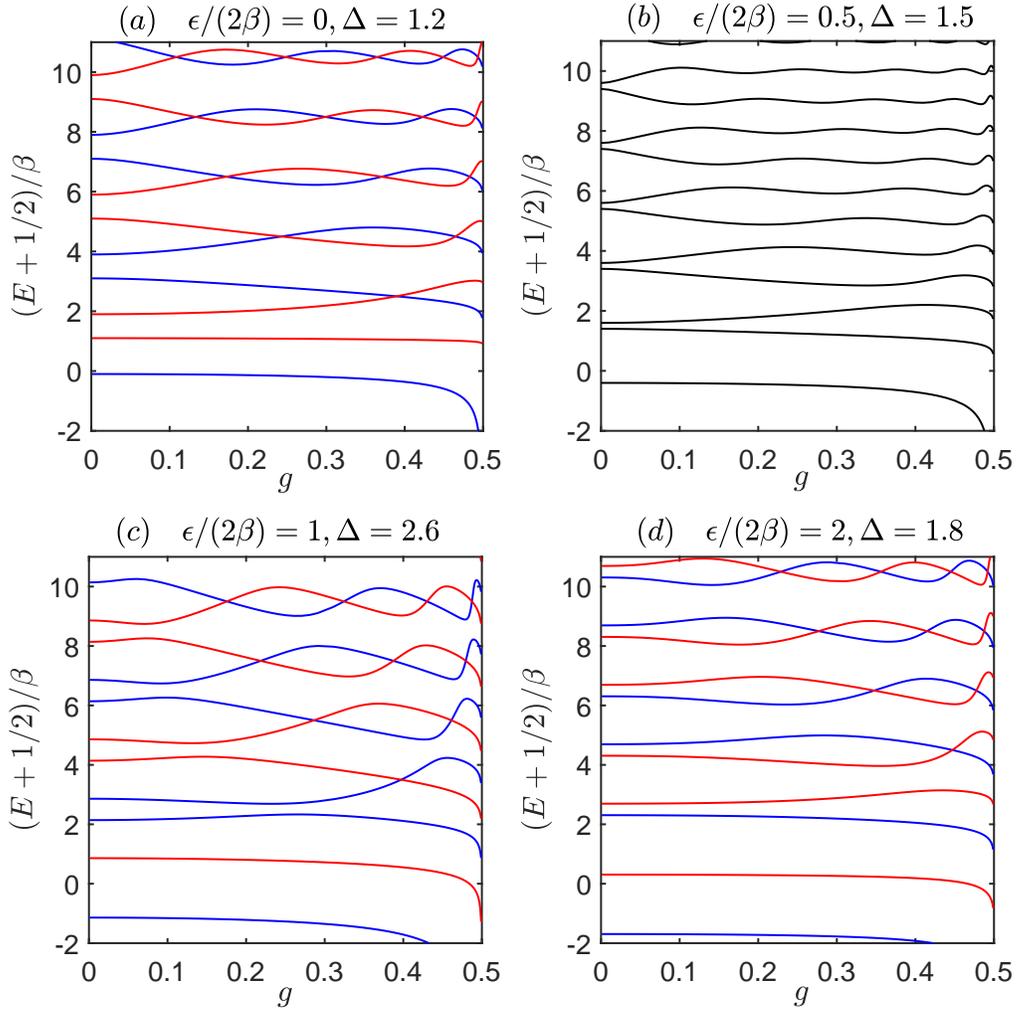}
\caption{(Color online) Energy spectrum as a function of the coupling
strength $g$ for (a) $\protect\epsilon/(2\protect\beta)=0$, (b) $\protect%
\epsilon/(2\protect\beta)=0.5$, (c) $\protect\epsilon/(2\protect\beta)=1$,
and (d) $\protect\epsilon/(2\protect\beta)=2$ in the $1/4$-subspace. The
values of $\Delta$ are randomly selected between $1$ and $3$. For clarity,
the energies are rescaled with $(E+1/2)/\protect\beta$. The blue (red) lines
correspond to the positive (negative) parity. }
\label{spectrum}
\end{figure}

According to the definition of the symmetry operator $J$ for the asymmetric
tpQRM, we have%
\begin{equation}
J^{2}=Q^{\dag }Q.
\end{equation}
For $N=0$, one can immediately obtain $J_{0}^{2}=I$ ($I$ is the $2\times 2$
unit matrix), which results in the eigenvalues $\pm 1$ of $J_{0}$. It
coincides with the usual definition of the parity operator for the symmetric
case \cite{duan2016}. For the asymmetric tpQRM, we find $\left( J_{1}\right)
^{2}$ is in a quadratic polynomial in $H_{1}$ following as
\begin{equation}
\left( J_{1}\right) ^{2}=\frac{1}{4\beta ^{2}}\left( H_{1}\right) ^{2}+\frac{%
1}{4\beta ^{2}}H_{1}+\left( \frac{\Delta ^{2}}{64g^{2}}+\frac{g^{2}}{4\beta
^{2}}\right) I.  \label{J1_sq}
\end{equation}%
Therefore, for the subspace, if an eigenvalue $E$ of the Hamiltonian $H_{1}$
is given, the operator $J_{1}$ has two possible eigenvalues $J_{1}(E)$, where%
\[
J_{1}(E)=\pm \sqrt{\frac{1}{4\beta ^{2}}E^{2}+\frac{1}{4\beta ^{2}}E+\frac{%
\Delta ^{2}}{64g^{2}}+\frac{g^{2}}{4\beta ^{2}}}.
\]%
The subspace will be again divided into two sectors with positive and
negative eigenvalues of $J_{1}$, which may still imply a $\mathbb{Z}_{4}$%
-symmetry in the asymmetric tpQRM although the separation depends on the
system parameters.

We also calculated $J$-square for $N=2,3$, resulting in%
\begin{equation}
J_{N}^{2}=\sum_{i=0}^{2N}y_{i}^{(N)}\left( g,\beta ,\Delta \right) \left(
H_{N}\right) ^{i},  \label{JN_sq}
\end{equation}%
where $y_{i}^{(N)}$ are the coefficients. Interestingly, the highest degree
of the polynomial (\ref{JN_sq}) is $2N$ in the asymmetry tpQRM rather than $%
N $ in the one-photon AQRM at the same order level crossings. This is
because the nolinear two-photon interaction, similar to the asymmetric
Rabi-Stark model case where the Stark term $\Delta a^{\dag }a\sigma _{z}/2$
also induce the $2N$-degree in the $J$-square \cite{Xilin2}, which need
further research.

To define a parity operator $\Pi_N$ independent of the system parameters, we
can rescale the symmetry operator, following as
\begin{equation}
\Pi_N = \frac{J_N}{\sqrt{\sum_{i=0}^{2N}y_{i}^{(N)}\left( H_{N}\right) ^{i}}}%
.  \label{pi}
\end{equation}
The eigenvalues of the parity $\Pi_N$ is $\pm 1$, analogous to the parity in
the symmetric case, i.e. $J_0$. The energy spectra of the $1/4$-subspace are
illustrated numerically in figure \ref{spectrum} for different $%
\epsilon/(2\beta)$ and $\Delta$, which is similar to the figure 7 in
reference \cite{xie2021}. In figure \ref{spectrum} (a),(c) and (d), for the
integer $\epsilon/(2\beta)=0,1,2$, the energy levels can be divided into two
sectors with positive and negative parities. The positive (negative) parity
is denoted by blue (red) lines, which can be achieved by solving the
eigenvalues of equation (\ref{pi}). Obviously, the level crossings between
the positive and negative parities are double degenerate which have been
solved through polynomials in reference \cite{xie2021}. As a comparison, in
figure \ref{spectrum} (b), where $\epsilon/(2\beta)=0.5$ breaks the
symmetry, the level crossings disappear in the spectrum and hereby the
symmetry operator is absent. Note the existence of level crossings is $\Delta
$ independent as long as $\Delta \neq 0$, shown in the figure 1 with random
values of $\Delta$. These results indicate that only when $%
\epsilon/(2\beta)=N$ with integer $N$ for the asymmetric tpQRM, the level
crossings can appear in the spectrum associated with the symmetry operator
which can label the energy levels as positive and negative parities.

\section{Conclusion}

In this work, we have developed a BOA scheme to derive the symmetry operator
responsible for the hidden symmetry of the asymmetric tpQRM systematically.
We demonstrate explicit solutions of the symmetry operators at low biases.
The solutions consist of the Lie algebra derived from the Bogoliubov
operators in a compact way, and the derivation is very concise and
accessible. For the asymmetric tpQRM, the Bogoliubov operator in our BOA
scheme is the linear combination of the original operators in the Fock
space, which means a few Bogoliubov operators could capture many more
original operators. It would sharply decrease the power series of the
solutions. Besides, the Lie algebra can reduce half degree of the original
operators. Therefore, the BOA scheme by the combination of Bogoliubov
operators and Lie algebra greatly increases the efficiency and clarity to
deal with the symmetry operators than the previous approach \cite%
{man,rey1,rey2} in the original Fock space. In addition, the square of the
symmetry operator for the asymmetric tpQRM can be expanded in a polynomial
of the Hamiltonian, showing twice order of that in the one-photon AQRM at
the same order level crossings because of the nolinear two-photon process.
For the asymmetric tpQRM, the energy levels in the spectrum can be divided
into positive and negative parities based on the corresponding symmetry
operator, and the level crossings reappear, similar to the symmetric case.
We believe the approach within BOA can be easily extended to the other more
complicated light-matter interaction systems to detect the hidden symmetry.

Finally, we would like to present some remarks. For the various one-photon
AQRMs, because the Bogoliubov operators are actually expressed linearly in
terms of the original operator, although the previous method based on the
expansions of the elements in $2\times 2$ matrix in the Fock space is rather
complicated, it still works. In contrast, for the nonlinear coupling
systems, such as the asymmetric two-photon QRM, it is unclear whether the
nature of the hidden symmetry can be reached by the direct expansion in the
Fock space. Interestingly, the BOA could be very effective both in the
analytical solutions and the attempt to find the underlying symmetry
operators.

Note added: The preliminary symmetry operators of the asymmetric tpQRM for
low biases $\epsilon =2\beta $ and $4\beta $ were given in our first version
of this arXiv preprint \cite{xie2021b}. Interestingly, the same idea
was used to derive the lowest symmetry operator in the asymmetric two-mode
quantum Rabi model most recently \cite{yan}.

\ack{The authors thank Liwei Duan for helpful discussions. This work was supported by the National Science
Foundation of China under Grant No. 11834005 and the National Key Research
and Development Program of China under Grant No. 2017YFA0303002.}


\end{document}